# Where and How is Entropy Generated in Solar Energy Conversion Systems?

Bolin Liao[1]*

[1]Department of Mechanical Engineering, University of California, Santa Barbara, CA, 93106, USA

## Abstract

The hotness of the sun and the coldness of the outer space are inexhaustible thermodynamic resources for human beings. From a thermodynamic point of view, any energy conversion systems that receive energy from the sun and/or dissipate energy to the universe are heat engines with photons as the "working fluid" and can be analyzed using the concept of entropy. While entropy analysis provides a particularly convenient way to understand the efficiency limits, it is typically taught in the context of thermodynamic cycles among quasi-equilibrium states and its generalization to solar energy conversion systems running in a continuous and non-equilibrium fashion is not straightforward. In this educational article, we present a few examples to illustrate how the concept of photon entropy, combined with the radiative transfer equation, can be used to analyze the local entropy generation processes and the efficiency limits of different solar energy conversion systems. We provide explicit calculations for the local and total entropy generation rates for simple emitters and absorbers, as well as photovoltaic cells, which can be readily reproduced by students. We further discuss the connection between the entropy generation and the device efficiency, particularly the exact spectral matching condition that is shared by infinite-junction photovoltaic cells and reversible thermoelectric materials to approach their theoretical efficiency limit.

* To whom correspondence should be addressed. Email: bliao@ucsb.edu

## I. Introduction

In the context of the radiative energy transfer, the Sun can be approximated as a blackbody at a temperature of 6000 K[1], and the outer space is filled with the cosmic microwave background at an effective temperature of 2.7 K. Any energy conversion devices on the Earth that extract useful work from the incoming solar radiation and/or emit radiation to the outer space can be analyzed thermodynamically[2] as heat engines. Indeed, numerous previous studies have proposed various efficiency limits[3–11] for these devices using thermodynamic arguments, as summarized by several excellent monographs and textbooks[12–15]. A key concept in these arguments is the entropy of radiation, which can be traced back to Planck[16] and was later discussed by Landau[17]. Although the entropy of radiation has the same thermodynamic meaning as the entropy of other working fluids (e.g. Carnot engines can be constructed[18] using a photon gas as the working fluid), it is not straightforward for students to understand the entropy generation and transport processes in radiative energy conversion devices that run in a continuous fashion. Despite the existence of extensive treatments of the thermodynamics of thermal radiation in the pedagogical literature[2,19–22], explicit calculations of the entropy generation *locally* in radiative energy conversion devices have not been provided. In this article, we present a general framework to calculate the local and total entropy generation rates in radiative energy conversion devices using the concept of photon entropy and the radiative transfer equation[23–25], and provide several examples to illustrate the entropy generation process in simple devices that can be taught in courses of thermodynamics and heat transfer.

At the thermal equilibrium, the energy and entropy densities of the blackbody radiation can be calculated by counting the number of available modes of electromagnetic (EM) waves and the photon occupation number for each mode[15,26]. In a macroscopic medium with a refractive index $n$, the number of available modes of EM waves per volume and per frequency interval is given by the optical density of states $D(\omega) = \frac{n^3 \omega^2}{\pi^2 c^3}$, where $\omega$ is the angular frequency and $c$ is the speed of light in the vacuum. At the thermal equilibrium with a temperature $T$, the number of photons $N(\omega)$ occupying each available mode is given by the Bose-Einstein distribution $N_0(\omega) = 1 \big/ \left( e^{\frac{\hbar\omega}{k_b T}} - 1 \right)$, where $\hbar$ is the reduced Planck constant, $k_b$ is the Boltzmann constant and $\hbar\omega$ is the energy of a photon. Therefore, the total energy density of the blackbody radiation at the thermal equilibrium can be calculated as

$$E_0 = \int_0^{+\infty} \hbar\omega N_0(\omega) D(\omega) d\omega, \tag{1}$$

and the Stefan-Boltzmann law of the blackbody emissive power can be derived from this integration[26].

If we treat the photons occupying each mode as molecules in a box of gas, the entropy associated with the blackbody radiation can be calculated using standard procedures in the equilibrium statistical mechanics[27]. Particularly, when the intensity of the blackbody radiation is low, the photons can be treated as independent bosons, due to the small cross section of photon-photon interactions. Under this condition, Rosen[28] derived the expression of entropy *at the thermal equilibrium* for an EM wave mode occupied by $N(\omega)$ independent photons:

$$s(\omega) = k_b[(1+N)\ln(1+N) - N\ln N]. \tag{2}$$

This dependence of the entropy on the photon occupation number is plotted in Fig. 1(a). At the thermal equilibrium, $N(\omega) = N_0(\omega)$, and the equilibrium value for the mode specific entropy is

$$s_0(\omega) = k_b[(1+N_0)\ln(1+N_0) - N_0\ln N_0] = N_0 \frac{\hbar\omega}{T} - k_b \ln\left(1 - e^{-\frac{\hbar\omega}{k_b T}}\right). \tag{3}$$

An integration over all available modes leads to the total entropy density of the blackbody radiation at the thermal equilibrium

$$S_0 = \int_0^{+\infty} s_0(\omega)D(\omega)d\omega$$

$$= \frac{1}{T}\int_0^{+\infty} \hbar\omega N_0(\omega)D(\omega)d\omega - \int_0^{+\infty} k_b \ln\left(1 - e^{-\frac{\hbar\omega}{k_bT}}\right)\frac{n^3\omega^2}{\pi^2c^3}d\omega$$

$$= \frac{E_0}{T} + \frac{n^3}{3\pi^2c^3T}\int_0^{+\infty} \hbar\omega^3 \frac{1}{e^{\frac{\hbar\omega}{k_bT}}-1}d\omega$$

$$= \frac{E_0}{T} + \frac{1}{3T}\int_0^{+\infty} \hbar\omega N_0(\omega)D(\omega)d\omega = \frac{4E_0}{3T},$$

(4)

where integration by parts is used from line 2 to line 3. The extra factor of $\frac{1}{3}$, as compared to the

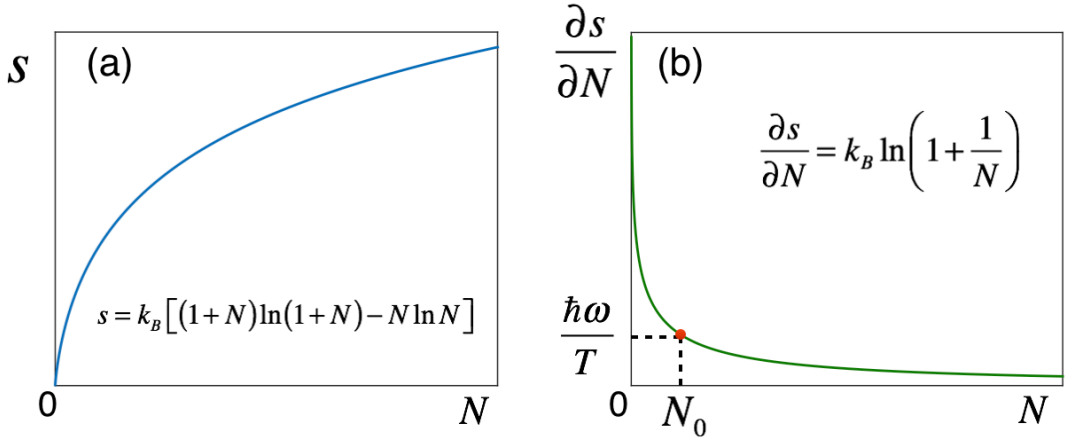

Figure 1 (a) The entropy of a photon mode as determined by the occupation number $N$ of the mode. (b) The change of the entropy of a photon mode as one photon is added or removed is plotted as a function of the photon occupation number $N$. At thermal equilibrium, when $N = N_0$, the change of entropy is given by $\frac{\hbar\omega}{T}$.

more familiar relation $S_0 = \frac{E_0}{T}$ for the energy exchange with a thermal reservoir, was first shown by Planck[16]. It creates a paradoxical scenario as pointed out by Würfel[24]. Suppose a black emitter is in thermal equilibrium with a thermal reservoir at a temperature $T$ and emits thermal radiation to the environment at the same time. Assume a heat flux $Q$ is received by the emitter from the reservoir, and the associated entropy flux is $\frac{Q}{T}$. At the steady state, the same amount of heat flux $Q$ must be emitted to the environment, carrying an outgoing entropy flux of $\frac{4}{3}\frac{Q}{T}$, indicating that a net amount of entropy is generated at a rate of $\frac{1}{3}\frac{Q}{T}$. At first sight, this result is surprising since everything is at the same temperature. The paradox can be resolved by recognizing that the emitter is, in fact, not in thermal equilibrium with the environment: implicitly, the environment is assumed to be at $T = 0$, as no photons are radiated back to the emitter[22,25].

To understand the details of entropy generation in the case above, it is necessary to generalize the entropy formula, Eq. 2, to non-equilibrium scenarios. Landsberg[29] demonstrated that Eq. 2 is also valid for non-equilibrium distributions of photons as long as the photons occupy the available modes *independently*. Similar derivations were also provided by other authors[19,30]. Indeed, the nonequilibrium photon entropy defined by Eq. 2 has been used as the basis to define the nonequilibrium temperature of photons[19,21]. Given Eq. 2 is also applicable to non-equilibrium photon distributions, it is instructive to calculate the entropy change when one photon is added to or removed from one mode, which is given by

$$\frac{\partial s(\omega)}{\partial N} = k_b \ln\left(1 + \frac{1}{N}\right) \tag{5}$$

and plotted in Fig. 1(b). When $N(\omega) = N_0(\omega)$, $\frac{\partial s(\omega)}{\partial N}$ is equal to $\frac{\hbar\omega}{T}$, implying that in this case if one photon is exchanged between the radiation field and a thermal reservoir at a temperature $T$, there is no net change of the entropy for the combined system. However, if $N(\omega) \neq N_0(\omega)$, the net entropy change for the combined system is nonzero when one photon is exchanged.

To further illustrate this observation, we can artificially divide the thermal emitter into two subsystems: the "material" consisting of atoms and electrons, and the radiation field (or "photon gas") contained within the emitter, as illustrated in Fig. 2. The two subsystems interact with each other by exchanging photons through the emission and absorption processes - photons can be absorbed and emitted by the atomic vibrations and electronic transitions in the material. While the emitter "material" is in a thermal equilibrium with a thermal reservoir at a temperature $T$, the local photon numbers of the radiation field are not necessarily equal to their equilibrium value $N_0(T, \omega)$. In particular, if $N(\omega) < N_0(T, \omega)$ $(\frac{\partial s(\omega)}{\partial N} > \frac{\hbar\omega}{T})$ and photons are emitted from the emitter material to the radiation field, or if $N(\omega) > N_0(T, \omega)$ $(\frac{\partial s(\omega)}{\partial N} < \frac{\hbar\omega}{T})$ and photons are absorbed by the emitter material from the radiation field, a net amount of entropy is generated in the combined system. These two scenarios can be clearly seen in Fig. 1(b). Therefore, to understand the entropy generation processes in an emitter or an absorber, it is crucial to compute the local distribution of photon numbers.

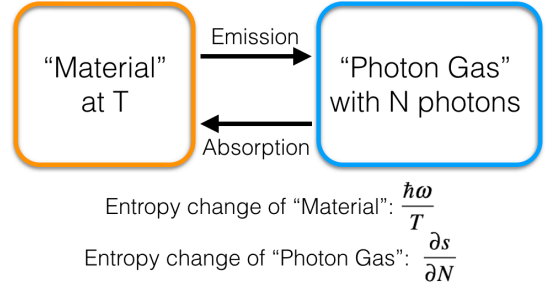

Figure 2 Schematic illustration of the two subsystems that coexist in an emitter: the "material" consisting of atoms and electrons and the "photon gas" that fills the emitter. Photons are exchanged between the two subsystems via emission and absorption, with corresponding entropy changes in both subsystems.

In radiative transfer, the local radiative intensity at the steady state can be calculated using the radiative transfer equation[27] (for simplicity, the optical scattering term is not included in the equation):

$$\frac{\partial I(\mathbf{r}, \theta, \phi, \omega)}{\partial \xi} = -\alpha I(\mathbf{r}, \theta, \phi, \omega) + \alpha I_b(T(\mathbf{r}), \omega), \qquad (6)$$

where $I(\mathbf{r}, \theta, \phi, \omega) = \frac{c}{4\pi n} \hbar\omega D(\omega) N(\mathbf{r}, \theta, \phi, \omega)$ is the spectral radiative intensity at a position $\mathbf{r}$ pointing to a direction specified by the polar angle $\theta$ and the azimuthal angle $\phi$, $I_b(T(\mathbf{r}), \omega) = \frac{c}{4\pi n} \hbar\omega D(\omega) N_0(T(\mathbf{r}), \omega)$ is the isotropic blackbody radiative intensity corresponding to the local temperature $T(\mathbf{r})$, $\xi$ is the optical path length and $\alpha$ is the (linear) absorption coefficient of the material. The two terms on the right hand side of Eq. 2 describe the local absorption and emission processes, respectively, where Kirchhoff's law equating the emissivity and the absorptivity is applied. By solving Eq. 6 with proper boundary conditions, the local photon number $N(\mathbf{r}, \theta, \phi, \omega)$ for each available EM wave mode can be computed, from which the local entropy density $s(\mathbf{r}, \theta, \phi, \omega)$ can be obtained using Eq. 2. We can further calculate the local entropy flux by integrating over all available modes:

$$\mathbf{J}_s = \iint s(\mathbf{r}, \theta, \phi, \omega) \frac{c}{n} \widehat{\mathbf{\Omega}} \frac{D(\omega)}{4\pi} \sin\theta \, d\theta d\phi d\omega, \tag{7}$$

where $\widehat{\mathbf{\Omega}}$ is a unit vector pointing to the direction $(\theta, \phi)$, and the integrations are over all directions and frequencies. In general, the number of photons occupying modes with the same frequency but traveling along different directions can be different due to boundary conditions and anisotropic radiative properties of the medium. Thus, the differential density of states $\frac{D(\omega)}{4\pi}$ is used and the integration over all directions is required in Eq. 7. Furthermore, the local entropy exchange $\dot{S}(\mathbf{r})$ between the radiation field and the material can be calculated using the entropy continuity equation at the steady state:

$$\dot{S}(\mathbf{r}) = \nabla \cdot \mathbf{J}_s = \iint \frac{\partial s(\mathbf{r}, \theta, \phi, \omega)}{\partial N(\mathbf{r}, \theta, \phi, \omega)} \nabla N(\mathbf{r}, \theta, \phi, \omega) \cdot \widehat{\mathbf{\Omega}} \frac{c}{n} \frac{D(\omega)}{4\pi} \sin\theta \, d\theta d\phi d\omega. \tag{8}$$

As the final step, the net local entropy generation rate $S_{gen}(\mathbf{r})$ of the combined system can be calculated as:

$$S_{gen}(\mathbf{r}) = \dot{S}(\mathbf{r}) - \frac{\dot{E}(\mathbf{r})}{T} = \iint \left( \frac{\partial s(\mathbf{r}, \theta, \phi, \omega)}{\partial N(\mathbf{r}, \theta, \phi, \omega)} - \frac{\hbar\omega}{T} \right) \nabla N(\mathbf{r}, \theta, \phi, \omega) \cdot \widehat{\mathbf{\Omega}} \frac{c}{n} \frac{D(\omega)}{4\pi} \sin\theta \, d\theta d\phi d\omega, \tag{9}$$

where $\dot{E}(\mathbf{r})$ is the local energy exchange rate between the radiation field and the "material". Combining Eqs. 6 and 9, noting that $\nabla N(\mathbf{r}, \theta, \phi, \omega) \cdot \widehat{\mathbf{\Omega}} = \frac{\partial N(\mathbf{r}, \theta, \phi, \omega)}{\partial \xi}$, leads to:

$$S_{gen}(\mathbf{r}) = \iint \alpha \left( \frac{\partial s(\mathbf{r},\theta,\phi,\omega)}{\partial N(\mathbf{r},\theta,\phi,\omega)} - \frac{\hbar\omega}{T} \right) \left[ N_0(T(\mathbf{r}),\omega) - N(\mathbf{r},\theta,\phi,\omega) \right] \frac{c}{n} \frac{D(\omega)}{4\pi} \sin\theta \, d\theta d\phi d\omega =$$

$$\iint \alpha \left[ \frac{k_b}{\hbar\omega} \ln\left(1 + \frac{1}{N}\right) - \frac{1}{T} \right] \left[ I_b(T(\mathbf{r}),\omega) - I(\mathbf{r},\theta,\phi,\omega) \right] \sin\theta \, d\theta d\phi d\omega. \tag{10}$$

Since $\left( \frac{\partial s(\mathbf{r},\theta,\phi,\omega)}{\partial N(\mathbf{r},\theta,\phi,\omega)} - \frac{\hbar\omega}{T} \right)$ and $\left[ N_0(T(\mathbf{r}),\omega) - N(\mathbf{r},\theta,\phi,\omega) \right]$ always have the same sign, Eq. 10 indicates that the local entropy generation is non-negative, consistent with the second law of thermodynamics. After $I(\mathbf{r},\theta,\phi,\omega)$ is solved using the radiative transfer equation (Eq. 6), Eq. 10 can be directly used to compute the local entropy generation rate during the radiative transfer. Equations 6 to 9 constitute the radiative entropy transfer equations and can be traced back to Planck[23]. Given the analogy between the phonon Boltzmann transport equation and the radiative transfer equation[31], Eq. 10 can also be used to study the entropy generation in the ballistic phonon transport[32].

## II. Entropy Generation during Emission and Absorption

In this section, we apply the framework presented in Section I to analyze two simple cases: black emitters and absorbers with a constant temperature $T$. For simplicity, we assume the refractive index $n = 1$ in the following discussion, therefore no additional treatment of the surface reflection and refraction is needed. We further assume that the emitter and the absorber considered here are optically thick, namely their thicknesses are much larger than the absorption depth $\frac{1}{\alpha}$. For optically thin materials, the above framework is still valid but the solution would be more complicated. In the simple emitter case, using the coordinate system shown in Fig. 3(a), Eq. 6 can be solved with the boundary conditions:

$$I(z = 0, \theta, \omega) = 0 \text{ for } 0 < \theta < \frac{\pi}{2} \text{ and } I(z = +\infty, \theta, \omega) = I_b(T, \omega) \text{ for } \frac{\pi}{2} < \theta < \pi \tag{11}$$

The boundary conditions reflect the fact that there is no radiation going from the environment into the emitter. The solutions are:

$$I(z, \theta, \omega) = I_b(T, \omega)\left(1 - e^{-\frac{\alpha z}{\cos\theta}}\right) \text{ for } 0 < \theta < \frac{\pi}{2} \tag{12}$$

$$I(z, \theta, \omega) = I_b(T, \omega) \text{ for } \frac{\pi}{2} < \theta < \pi,$$

as plotted in Fig. 3(b) with different values of $\theta$. Due to the deficiency of the incoming radiation flux, the average photon number near the emitter surface is below the thermal equilibrium value. In this case, as a net photon flux is emitted from the material to the radiation field, a net amount of entropy is generated, as discussed in the previous section[25]. The local entropy generation rate can be calculated using Eq. 10 and is plotted in Fig. 3(c). Most of the entropy is generated within one absorption depth $\frac{1}{\alpha}$ from the surface of the emitter, where the deviation of the photon occupation number from its equilibrium value is the largest. The total entropy generation rate can be calculated analytically[25] by integrating the local entropy generation rate over space to be $\frac{1}{3}\frac{Q}{T}$, where $Q$ is the total emissive power from the emitter surface. Indeed, when an emitter with a constant temperature $T$ emits to the environment with a net heat flux $Q$, the entropy is generated at a rate of $\frac{1}{3}\frac{Q}{T}$ due to the non-equilibrium distribution of photons near the emitter surface. If, however, the environment is at the same temperature $T$ and emits back to the emitter, then the radiation intensity inside the emitter would be $I_b(T)$ everywhere and there would not be generation of the entropy. Therefore, the generation of entropy is not intrinsically associated with the emission process itself, but depends on whether the emitter is at thermal equilibrium with the environment.

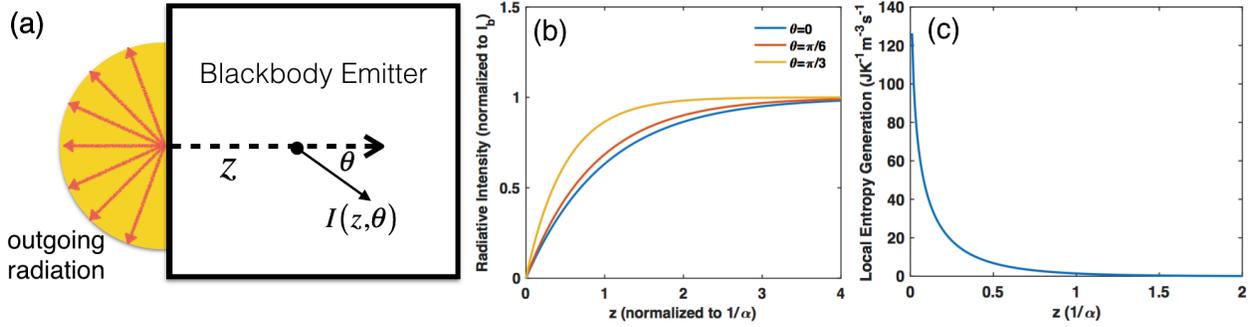

Figure 3 Local entropy generation in a simple blackbody emitter. (a) Schematic of the blackbody emitter. (b) The radiative intensities of modes traveling towards the inside of the emitter. The outgoing modes have radiative intensities of $I_b(T, \omega)$. (c) The local entropy generation rate inside the emitter. Here an absorption depth $\frac{1}{\alpha} = 1\ \mu m$, and temperature $T = 300\ K$ are assumed. The major contribution to the entropy generation is from the region close to the emitter surface.

Similarly, we can calculate the entropy generation rate when a black absorber at a temperature $T$ absorbs the thermal radiation from a source at a higher temperature $T_s > T$, as illustrated in Fig. 4(a). In this case, the boundary condition for the thermal radiation going into the absorber becomes $I(z = 0, \theta, \omega) = I_b(T_s, \omega)$ for $0 < \theta < \frac{\pi}{2}$ and the solution becomes

$$I(z, \theta, \omega) = I_b(T_s, \omega)e^{-\frac{\alpha z}{\cos\theta}} + I_b(T, \omega)\left(1 - e^{-\frac{\alpha z}{\cos\theta}}\right) \text{ for } 0 < \theta < \frac{\pi}{2}, \qquad (13)$$

while the solution for the outgoing radiation stays the same as the emitter case. Assuming $I_b(T_s, \omega) = 5I_b(T, \omega)$ for a certain frequency, the solutions are plotted in Fig. 4(b), showing that the photon occupation number near the absorber surface is above the equilibrium value at the temperature $T$ due to the higher temperature of the incoming radiation. As a net flux of photons is absorbed in this region, a net amount of entropy is generated, as shown in Fig. 4(c), where $T_s = 6000$ K is used for the case of the solar radiation. Due to the large difference between $T_s$ and $T$, a significant local entropy generation extends to three times the absorption depth into the absorber. Integrating the local entropy generation rate over space leads to the total entropy generation rate

$$S_{gen,total} = \frac{1}{3}\frac{Q_{out}}{T} + \left(\frac{T_s}{T} - \frac{4}{3}\right)\frac{Q_{in}}{T_s}, \qquad (14)$$

where $Q_{out}$ is the energy flux emitted by the blackbody absorber and $Q_{in}$ is the energy flux associated with the incoming radiation. This result was also first discussed by Planck[16]. When the incoming radiation has a much higher temperature than that of the absorber, namely $T_s \gg T$, as is the case for the solar absorbers, the total entropy generation is approximately $\frac{T_s}{T}S_{in} \gg S_{in}$, where $S_{in}$ is the incoming entropy flux. Therefore, in solar thermal energy converters, where the solar radiation is directly used to heat up an absorber, the main source of entropy is due to the entropy generation caused by the absorption process instead of the entropy input associated with the solar radiation. From the above discussion, it is clear that to suppress the entropy generation associated with the absorption process, one needs to boost the emission from the absorber $I_b$ to match the incoming $I_b(T_s, \omega)$. This can be achieved by either increasing the temperature of the absorber or making use of the chemical potential of the emitted photons[24], such as in photovoltaic cells, which will be discussed in the next Section.

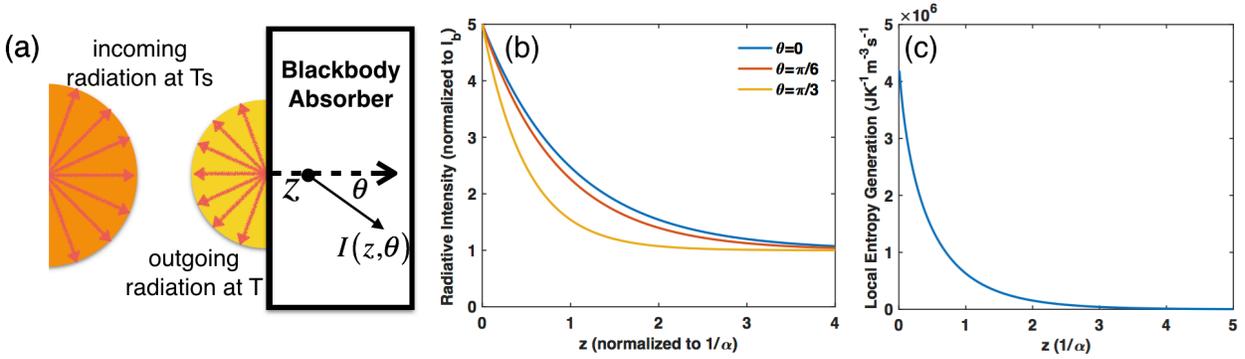

Figure 4 Local entropy generation for a simple absorber. (a) Schematic showing the simple blackbody absorber at temperature $T$ receiving radiation at a higher temperature $T_s$. (b) The radiative intensities of modes traveling towards the inside of the absorber, assuming $I_b(T_s, \omega) = 5I_b(T, \omega)$. The outgoing modes have radiative intensities of $I_b(T, \omega)$. (c) The local entropy generation in the absorber, assuming $\frac{1}{\alpha} = 1\ \mu m$, $T_s = 6000\ K$ and $T = 300\ K$.

The simple entropy analysis given above can be applied to understand the efficiency limits of solar energy conversion systems. In the absorber case, assuming the photon occupation number near the absorber surface can be controlled to completely suppress the entropy generation, one can calculate the maximum power output from such a system allowed by the second law of thermodynamics. To balance the entropy flow in the absorber, the difference of the incoming radiative entropy flux $S_{in} = \frac{4}{3}\sigma T_s^3$ and the outgoing radiative entropy flux $S_{out} = \frac{4}{3}\sigma T^3$ can be carried away by the thermal conduction. The amount of energy required to dissipate this amount of entropy is $T(S_{in} - S_{out}) = \frac{4}{3}\sigma(T_s^3 - T^3)T$, so the work extractable from the system is $\sigma(T_s^4 - T^4) - \frac{4}{3}\sigma(T_s^3 - T^3)T$, leading to the maximum efficiency with zero entropy generation

$$\eta_l = \frac{\sigma(T_s^4 - T^4) - \frac{4}{3}\sigma(T_s^3 - T^3)T}{\sigma T_s^4} = 1 - \frac{4}{3}\frac{T}{T_s} + \frac{1}{3}\frac{T^4}{T_s^4}, \tag{15}$$

known as the Landsberg efficiency[5]. Given $T_s = 6000\ \text{K}$ and $T = 300\ \text{K}$, the Landsberg efficiency is about 93%, lower than the Carnot efficiency $1 - \frac{T}{T_s} =$95%. The difference between the Landsberg efficiency and the Carnot efficiency is due to the fact that, in the solar energy conversion system analyzed above, the input entropy is purely carried by the thermal radiation, but the output entropy is partly carried by the thermal conduction, whose entropy to energy ratio ($\frac{1}{T}$) is lower than that of the thermal radiation ($\frac{4}{3}\frac{1}{T}$). Hsu et al. analyzed[33] this perspective systematically. This view is also consistent with the efforts of achieving a more efficient photon extraction from photovoltaic cells to boost their efficiency[34]. From an entropic point of view, it is beneficial to dissipate the heat from photovoltaic cells by thermal radiation instead of conduction.

Another known efficiency limit is the so-called "blackbody limit"[12,13], e.g. for concentrated solar power systems[35]. In this case, a blackbody absorber is used to receive the solar radiation and reaches a high temperature, and the difference between the energy fluxes associated with the incoming solar radiation and the outgoing radiation from the absorber is converted by a Carnot engine to work. It is found that the optimum absorber temperature in such a system is about 2500 K, corresponding to an optimum efficiency of about 85%. This efficiency is lower than the Landsberg efficiency, which can also be understood from the entropic point of view: whereas the entropy fluxes are balanced at the absorber for the Landsberg efficiency, the energy fluxes are balanced at the absorber for the blackbody limit. An entropy analysis of the absorber in the case of the blackbody limit reveals that a net amount of entropy is generated in the absorber even at the optimum absorber temperature, thus limiting the efficiency to be lower than the Landsberg efficiency.

## III. Entropy Analysis of Photovoltaic Cells

The discussion in the previous Section shows that to achieve a higher efficiency in solar energy conversion systems, it is important to suppress the entropy generation associated with the absorption of the solar radiation, which dominates the incoming entropy carried by the solar radiation due to the very high temperature of the Sun. For a simple black absorber, the only way is to increase the temperature of the absorber by concentrating the solar radiation and minimizing the heat loss[35]. There are ongoing efforts in developing high temperature compatible infrastructure[36] to boost the efficiency of concentrated solar power plants. For photovoltaic (PV) cells, where the "absorber" is a semiconductor with a band gap, there is an extra knob to tune the photon occupation numbers inside the absorber - the output voltage $V$ of the PV cell. When a PV cell is operating under the Sun, the incoming photons excite excess electrons and holes beyond their equilibrium densities, balanced by recombination events and the charge extraction to the external circuit[3]. The excess electron and hole densities are characterized by quasi Fermi-Dirac distributions with their quasi Fermi levels $\mu_e$ and $\mu_h$, respectively, and the output voltage of the solar cell $V$ is determined by the difference of the quasi Fermi levels $qV = \mu_e - \mu_h$, where $q = 1.6 \times 10^{-19}$ C is the charge of an electron[37]. At the steady state, the photons emitted by the quasi-equilibrium distributions of electrons and holes through the radiative recombination approach a quasi-equilibrium Bose-Einstein distribution with a chemical potential[20,24,38,39] determined by the output voltage:

$$\widetilde{N_0}(T,V,\omega) = 1 \bigg/ \left( e^{\frac{\hbar\omega - qV}{k_b T}} - 1 \right),$$

(16)

Therefore, the output voltage can control the photon occupation number inside the solar cell, in addition to the operating temperature, to affect the entropy generation process.

Here we analyze an ideal single-junction solar cell operating at the Shockley-Queisser conditions[3]: 1) each photon with an energy above the band gap is absorbed and generates one electron-hole pair while photons with energy below the band gap are not absorbed; 2) the radiative recombination is the only recombination channel for excess electrons and holes. Non-radiative recombination processes always lead to additional entropy generation and should be avoided as much as possible for a higher efficiency. With these assumptions, the framework presented in Section I can be readily applied here with the following modifications: 1) only photons with an energy above the band gap of the solar cell need to be considered; 2) the local blackbody radiation intensity $I_b(T, \omega)$ is replaced by

$$\widetilde{I_b}(T(\mathbf{r}), V(\mathbf{r}), \omega) = \frac{c}{4\pi n} \hbar \omega D(\omega) \widetilde{N_0}(T(\mathbf{r}), V(\mathbf{r}), \omega) = \frac{c}{4\pi n} \hbar \omega D(\omega) \left( e^{\frac{\hbar\omega - qV}{k_b T}} - 1 \right)^{-1}, \quad (17)$$

in Eq. 6; 3) when a photon is emitted or absorbed by the "material" at a temperature $T$ and a voltage bias $V$, the entropy change of the "material" is $\frac{\hbar\omega - qV}{T}$, instead of $\frac{\hbar\omega}{T}$. With these modifications, the local entropy generation rate inside a PV cell at a given temperature and voltage can be calculated following a similar procedure as in Section II. When the PV cell is forward biased, namely $V$ is positive, the emissive power from the PV cell is increased above the value of the blackbody radiation at the same temperature, and thus the entropy generation due to the absorption of the solar radiation is reduced. We focus on the total entropy generation rate as a means to understand various efficiency limits of a PV cell. The energy, photon number and entropy fluxes associated with the incoming solar radiation and the outgoing radiation emitted by the PV cell, as well as the energy and entropy fluxes dissipated by the thermal conduction are listed in Table I[28]. Here $h$ is the Planck constant and $\omega_s = 6.87 \times 10^{-5}$ sr is the solid angle of the Sun seen from the Earth, corresponding to the solar angle $\theta_s = 9.35 \times 10^{-3}$ radian. The difference between the incoming and the outgoing photon number fluxes is equal to the number of electrons extracted from the PV cell[3], generating an output electrical power $P$.

Table I. The energy, photon number and entropy fluxes associated with the incoming solar radiation and the outgoing emission from a PV cell. The difference between the energy fluxes is dissipated by thermal conduction, while the difference between the photon number fluxes leads to an output charge current.

| | Incoming Solar Radiation | Outgoing Radiation | Heat Dissipated by Conduction |
|---|---|---|---|
| Energy Flux | $Q_i = \omega_s \dfrac{2(k_b T_s)^4}{h^3 c^2} \displaystyle\int_{x_g}^{+\infty} \dfrac{x^3 dx}{e^x - 1}$ | $Q_{oe} = \dfrac{2\pi(k_b T_c)^4}{h^3 c^2} \displaystyle\int_{x_g/x_c}^{+\infty} \dfrac{x^3 dx}{e^{x-\gamma} - 1}$ | $Q_{oc} = Q_i - Q_{oe} - P$ |
| Photon Number Flux | $A_i = \omega_s \dfrac{2(k_b T_s)^3}{h^3 c^2} \displaystyle\int_{x_g}^{+\infty} \dfrac{x^2 dx}{e^x - 1}$ | $A_{oe} = \dfrac{2\pi k_b^3 T_c^3}{h^3 c^2} \displaystyle\int_{x_g/x_c}^{+\infty} \dfrac{x^2 dx}{e^{x-\gamma} - 1}$ | |
| Entropy Flux | $S_i = \omega_s \dfrac{2k_b^4 T_s^3}{h^3 c^2} \displaystyle\int_{x_g}^{+\infty} f(x) x^2 dx$ | $S_{oe} = \dfrac{2\pi k_b^4 T_s^3}{h^3 c^2} \displaystyle\int_{x_g/x_c}^{+\infty} f(x-\gamma) x^2 dx$ | $S_{oc} = \dfrac{Q_i - Q_{oe} - P}{T}$ |
| Power Output | $P = q(A_i - A_{oe})V$ | | |
| $x_g = \dfrac{E_g}{k_b T_s}, \quad x_c = \dfrac{T}{T_s}, \quad \gamma = \dfrac{qV}{k_b T_c}, \quad f(x) = \left(1 + \dfrac{1}{e^x - 1}\right)\ln\left(1 + \dfrac{1}{e^x - 1}\right) - \dfrac{1}{e^x - 1}\ln\dfrac{1}{e^x - 1}$ | | | |

The total entropy generation in the PV cell as a function of the output voltage is shown in Fig. 5(a). Here the band gap of silicon $E_g = 1.1$ eV is used. From Fig. 5(a), the entropy generation is at the minimum when $V = V_{oc}$, the open-circuit voltage. The source of the entropy generation can be divided into two parts: within the solar angle $0 < \theta < \theta_s$, the entropy generation is due to the absorption of the solar radiation, whereas outside the solar angle, the entropy generation is due to the emission of the PV cell. Figure 5(a) shows that when the output voltage is below the open-circuit voltage, including the optimum operating point $V_{op}$ corresponding to the maximum power output and the highest efficiency (the Shockley-Queisser limit[3], ~30% for this band gap), the major entropy generation is associated with the absorption process. When the voltage is above the open-circuit voltage, a net electrical power is put into the PV cell and the emitted photon number flux is higher than the incoming solar radiation. In this less useful scenario, the entropy generation is dominated by the emission process. The entropy generation associated with the emission process can be largely suppressed by limiting the emission of the PV cell within the solar solid angle, by using either an optical cavity[40] or an angle-selective emitting surface[41]. In the normal operating region of a PV cell ($V < V_{oc}$), the entropy generation associated with the absorption of the solar radiation is caused by the spectral mismatch of the incoming and outgoing radiation. For example, under the open circuit condition, although the total incoming and the outgoing photon number fluxes are balanced such that the output electrical current is zero, at each frequency the photon number fluxes are not equal, as shown in Fig. 5(b). While the emission from the PV cell is mostly concentrated in the region right above the band gap, the incoming solar radiation extends more into the higher energy region. Stacking PV cells with different band gaps is a well known way to balance the incoming and the outgoing radiation spectra, since the output voltage of each PV cell can be independently adjusted to match the solar spectra within a certain range. At the infinite junction limit (or "multi-color" limit[12]), where an infinite number of PV cells with different band gaps are stacked to match all frequencies within the solar spectrum, each individual frequency of the incoming solar radiation interacts with a separate PV cell. Therefore, under the open circuit condition, the incoming and the outgoing photon number fluxes are balanced at each frequency, leading to the corresponding open circuit voltages determined by

$$1 \Big/ \left( e^{\frac{\hbar\omega}{k_b T_s}} - 1 \right) = 1 \Big/ \left( e^{\frac{\hbar\omega - qV_{oc}}{k_b T}} - 1 \right) \tag{18}$$

$$V_{oc}(\omega) = \left( 1 - \frac{T}{T_s} \right) \frac{\hbar\omega}{q} \tag{19}$$

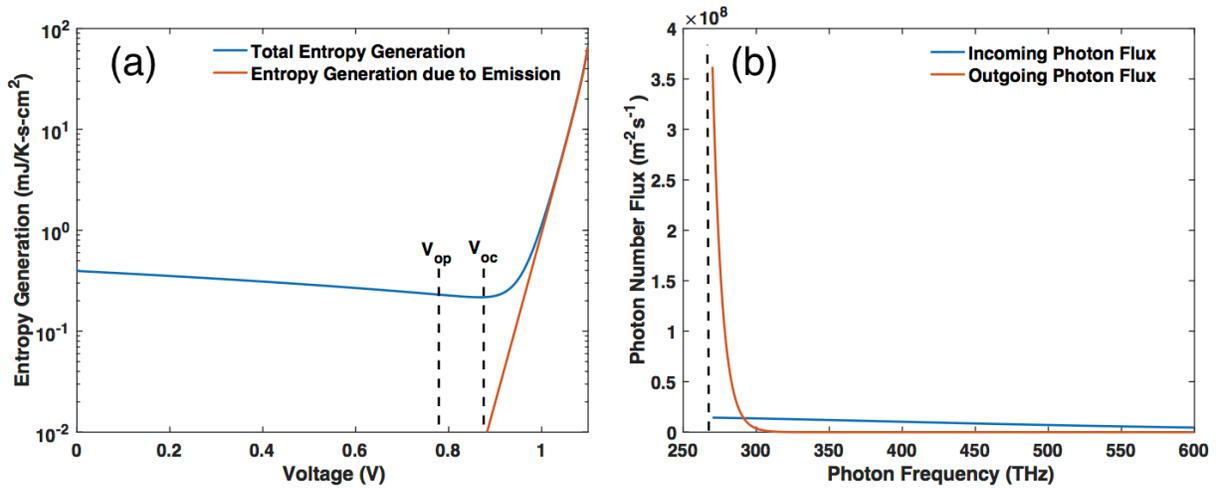

Figure 5 Entropy analysis for an ideal silicon PV cell with band gap 1.1 eV. (a) The total entropy generation in the PV cell as a function of its operating voltage. Also plotted is the entropy generation due to emission of the PV cell, while the rest of the total entropy generation is due to absorption. $V_{oc}$ and $V_{op}$ denote the open circuit voltage and the voltage for maximum power output, respectively. (b) The spectral photon number flux for incoming solar radiation and the radiation emitted by the PV cell at the open circuit voltage and temperature $T = 300 \ K$.

Due to this exact spectral matching, the entropy generation is zero for an infinite-junction PV cell at the open circuit voltage[12]. However, under the optimum power output condition, there is still a finite amount of entropy generation in an infinite-junction PV cell, whose theoretical efficiency is 86.8%, below the Landsberg efficiency that assumes zero entropy generation. It is interesting to note that the exact spectral matching condition of the incoming and the outgoing energy fluxes also leads to the theoretical optimum operation of other continuously operating energy conversion systems, such as thermoelectric devices[42]. Thermoelectric devices use electrons, instead of photons, as the "working fluid" to interconvert thermal and electrical energy[43]. Humphrey and Linke[42] showed that if the electrical energy in a thermoelectric material is only carried by electrons with a single energy level, then under the open circuit condition, the electron occupation number of this single energy level can be balanced everywhere in the material with different local temperatures by developing an electrical voltage such that $1 \Big/ \left( e^{\frac{E - qV(\mathbf{r})}{k_b T(\mathbf{r})}} + 1 \right) = $ constant, where $E$ is the energy of the single electron level. Under this condition, the entropy generation is zero despite that the local temperature in the material can be different everywhere. This so-called "reversible thermoelectrics" provide a thermodynamic explanation for the proposal by Mahan and Sofo[44] that the best thermoelectric materials should have a single electron energy level (or an electron density of states taking a Dirac delta-function form, as originally stated in their paper[44]). This connection between infinite-junction PV cells and reversible thermoelectrics points to a common strategy to boost the efficiency of continuously running "particle-exchange heat engines"[13,45].

## IV. Conclusion

In this article, a general framework for the radiative entropy analysis is summarized and applied to explicitly calculate the local and the total entropy generation in three cases: simple blackbody emitters, absorbers and ideal PV cells. These examples demonstrate that energy conversion devices using photons as the working fluid can be analyzed from the entropic point of view, just as heat engines and energy conversion systems based on thermodynamic cycles with liquid or gas as the working fluid. The framework presented in this article can be readily applied to analyze other solar energy conversion systems, such as concentrated solar thermal plants[35], thermoradiative cells[46], radiative cooling devices[47] and hybrid systems[14,48], and to include non-ideal effects. Systematically identifying the entropy generation mechanisms and devising ways to suppress the entropy generation channels can lead to innovative designs for radiative energy conversion systems with an improved efficiency.

**Acknowledgement**

The author acknowledges helpful discussions with Gang Chen, Svetlana Boriskina, Wei-Chun Hsu, Jonathan Tong, Vazrik Chiloyan, Lee Weinstein, Jiawei Zhou and Alejandro Vega-Flick. This work is supported by a startup fund from University of California, Santa Barbara.